\documentclass[preprint,showpacs,preprintnumbers,amsmath,amssymb]{revtex4-1}
\usepackage{graphicx,textcomp}
\usepackage{dcolumn}
\usepackage{bm}
\usepackage{amsmath}
\usepackage{color}

\begin{document}

\title{Photoassociation of a cold atom-molecule pair II: second order perturbation approach}
\author{M. Lepers$^{1}$, R. Vexiau$^{1}$,  N. Bouloufa$^{1}$, O. Dulieu$^{1}$, V. Kokoouline$^{1,2}$}
\affiliation{$^{1}$Laboratoire Aim\'e Cotton, CNRS, B\^at. 505, Universit\'e Paris-Sud, 91405 Orsay Cedex, France\\
$^{2}$Department of Physics, University of Central Florida, Orlando, Florida 32816, USA }
\email[M. Lepers: ]{maxence.lepers@u-psud.fr}

\begin{abstract}
The electrostatic interaction between an excited atom and a diatomic ground state molecule in an arbitrary rovibrational level at large mutual separations is investigated with a general second-order perturbation theory, in the perspective of modeling the photoassociation between cold atoms and molecules. We find that the combination of quadrupole-quadrupole and van der Waals interactions compete with the rotational energy of the dimer, limiting the range of validity of the perturbative approach to distances larger than 100 a.u.. Numerical results are given for the long-range interaction between Cs and Cs$_2$, showing that the photoassociation is probably efficient whatever the Cs$_2$ rotational energy.

\end{abstract}


\maketitle

\section{Introduction}

In a recent paper (hereafter referred to as paper I) \cite{lepers2010} we investigated the electrostatic interaction between an atom and a diatomic molecule at large mutual separations. This kind of study is particularly relevant in the context of the amazing development of researches on ultracold quantum gases, \textit{i.e.} when the kinetic energy of the relative motion of the particles inside the gas is equivalent to a temperature far below 1~Kelvin. Many enlightening review papers are available in the scientific literature, like the most recent ones \cite{carr2009,dulieu2009,chin2010} devoted to ultracold molecular gases which is our main purpose. Motivations and applications of such researches are thoroughly discussed in these reviews, and extensive bibliography can be found there that we will not repeat in the present paper. Briefly, dilute atomic and molecular gases at ultracold temperatures exhibit pure quantum effects, as their dynamics is sensitive to quantum resonances and quantum statistics leading to Bose or Fermi degeneracy. Ultracold gases exhibit unique physical conditions for high precision measurements related to the limit of validity of fundamental theories like the Standard Model or for quantum simulation of condensed matter phases like superfluidity or superconductivity.

At such a low energy regime, the relative motion of the particles inside the gas is controlled by their weak mutual interactions at large distance $R$, primarily induced by electrostatic forces described by a multipolar expansion in $R^{-n}$ terms. For instance, it is well known that spin-free or rotation-less particles in their ground state interact mainly through van der Waals potentials varying as $C_6/R^6$ resulting from a second-order perturbation treatment of the dipole-dipole interaction. Calculations of the long-range dispersion coefficients $C_6$ for alkali-metal or alkaline-earth atoms -which represent the species of choice for ultracold gases- have nowadays reached an unprecedented accuracy \cite{mitroy2003,derevianko2010} and represent an invaluable input for interpreting experiments. First-order terms may become dominant when one atom or both atoms are excited, resulting into interactions varying as $R^{-3}$ \cite{bouloufa2009} or $R^{-5}$ \cite{marinescu1997} for dipole-dipole and quadrupole-quadrupole interactions, respectively.

In principle, such calculations can be easily extended to the interaction between atoms and molecules with internal rotation. However most previously published studies have been restricted to the situation where molecules are fixed in space \cite{bussery-honvault2008,bussery-honvault2009}. A recent study actually treated the van der Waals interaction between two identical ground state molecules in a given rotational level \cite{kotochigova2010} using the second-order perturbation theory. In paper I, we were interested in the interaction between an excited atom and a ground state molecule in a rotational level, governed by a first-order quadrupole-quadrupole term varying as $R^{-5}$, that we evaluated through first-order degenerate perturbation theory. We demonstrated that this term competes with the rotational energy of the dimer, so that avoided crossings, and possibly long-range wells, could be expected in the long-range potential curves of the atom-dimer complex. Such patterns are relevant in the perspective of future studies aiming at associating an ultracold atom-molecule pair using laser light (photoassociation, or PA) to create stable ultracold triatomic molecules, according for instance to the reaction
\begin{equation}
\textrm{Cs}(6P)+\textrm{Cs}_2(X^1\Sigma_g^+) \rightarrow \textrm{Cs}_3^* \rightarrow \textrm{Cs}_3 + \textrm{photon} \,.
\label{eq:PA}
\end{equation}
In the present work, we extend our treatment to the second order of the perturbation development, to determine the next term varying as $R^{-6}$ which is expected to compete with the $R^{-5}$ term when $R$ decreases. As in paper I, we illustrate our development for the case of an alkali-metal ground state $X^1\Sigma_g^+$ molecule in an arbitrary rotational level $j$ with an  alkali-metal atom excited to the $n^2P$ electronic level, but it can be easily generalized to arbitrary species. In Section \ref{sec:c6} we first recall the general expression for the long-range multipolar expansion and the expression of the $C_6$ coefficients in the case of the van der Waals interaction between an atom and a molecule, and we relate these quantities to the dipole polarizabilities in imaginary frequencies, which are evaluated in Section \ref{sec:polar}. Potential curves for the long-range interaction between an excited Cesium atom and a ground state Cs$_2$ molecule are displayed in Section \ref{sec:results} before concluding remarks and perspectives (Section \ref{sec:conclusion}). Atomic units (a.u.) for distances (1 a.u. = 0.0529177~nm) and for energies (1~a.u. = 219474.63137~cm$^{-1}$) will be used throughout the paper, except otherwise stated.

\section{Expression of the van der Waals interaction}
\label{sec:c6}

\begin{figure}
\begin{centering}
\includegraphics[width=0.5\textwidth]{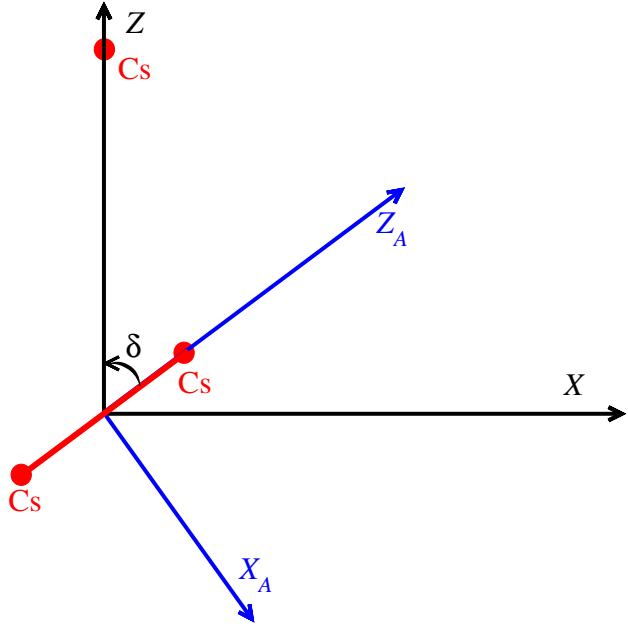}
\par\end{centering}
\caption{\label{fig:CS}The dimer $\left[X_AY_AZ_A\right]$ (D-CS), and $\left[XYZ\right]$ (T-CS) coordinate systems, defined for the dimer and for the trimer, respectively. The $Z_A$ axis is along the dimer axis, while $Z$ is oriented from the center of mass of the dimer towards the atom $B$. The $Y$ and $Y_A$ axes coincide and point into the plane of the figure. The T-CS is deduced from the D-CS by a rotation with an angle $\delta$ around the $Y$ axis. The Euler angles $(\alpha,\beta,\gamma)$ are not represented here.}
\end{figure}

We first recall briefly the notation used in paper I. The electrostatic potential energy between two charge distributions A (the dimer) and B (the atom) at large distance $R$ (\textit{i.e.} beyond the LeRoy radius \cite{leroy1974}) is expressed as the usual multipolar expansion
\begin{eqnarray}
\hat{V}_{AB}(R) & = & \sum_{L_{A},L_{B}=0}^{+\infty}\sum_{M=-L_{<}}^{L_{<}}\frac{1}{R^{1+L_{A}+L_{B}}}\nonumber \\
 & \times & f_{L_{A}L_{B}M}\hat{Q}_{L_{A}}^{M}(\hat{r}_{A})\hat{Q}_{L_{B}}^{-M}(\hat{r}_{B})\,,
\label{eq:LR-Potential}
\end{eqnarray}
where $\hat{Q}_{L_{X}}^{M}(\hat{r}_{X})$ is the operator associated with the $2^{L_{X}}$-pole of charge distribution $X$ $(X=A$ or $B)$, expressed in the coordinate system with the origin at the center of mass of $X$
\begin{equation}
\hat{Q}_{L_{X}}^{M}(\hat{r}_{X})=\sqrt{\frac{4\pi}{2L_{X}+1}}\sum_{i\in X}q_{i}\hat{r}_{i}^{L_{X}}Y_{L_{X}}^{M}\left(\hat{\theta}_{i},\hat{\phi}_{i}\right)\,,
\label{eq:multipole}
\end{equation}
where
\begin{eqnarray}
f_{L_{A}L_{B}M} & = & \frac{\left(-1\right)^{L_{B}}\left(L_{A}+L_{B}\right)!}{\sqrt{\left(L_{A}+M\right)!\left(L_{A}-M\right)!}}\nonumber \\
 & \times & \frac{1}{\sqrt{\left(L_{B}+M\right)!\left(L_{B}-M\right)!}}\label{eq:LR-gM-LALB}
\label{eq:LR-factor}
\end{eqnarray}
and $L_{<}$ is the minimum of $L_{A}$ and $L_{B}$. The $Z$ quantization axis for the projections $\pm M$ in the above equations is oriented from A to B, yielding the $\left(-1\right)^{L_{B}}$ factor in Eq. (\ref{eq:LR-factor}). We define two body-fixed (BF) coordinate systems (CS) displayed in Fig. \ref{fig:CS}: the dimer CS (or D-CS) with axes $X_A,Y_A,Z_A$, and the  trimer CS (or T-CS) with axes $X,Y,Z$. The T-CS is deduced from the D-CS by a rotation with an angle $\delta$ around the $Y$ axis. The T-CS is related to the space-fixed (SF) coordinate system $(\tilde{x} \tilde{y} \tilde{z})$ by the usual Euler angles $(\alpha,\beta,\gamma)$.

The dimer is assumed in an arbitrary vibrational $\left|v_d\right\rangle$ and rotational state $\left|j\right\rangle$  of its electronic ground state $\left|X^{1}\Sigma_{g}^{+}\right\rangle$. The atom $B$ is chosen with a single outer electron being excited to the $p$ state $\left|n,\ell=1\right\rangle $. The projections $m_j$ and $\lambda$ of $j$ and $\ell$ are defined with respect to the $Z$ axis. The energy origin corresponds to an infinite separation between the atom and the dimer. The first-order perturbation theory yields the quadrupole-quadrupole energies with zeroth-order wave functions that can be written as a linear combination of the various $\left|m_{j},\lambda\right\rangle$ substates of the complex
\begin{equation}
\left|\Phi_{0}^{0}\right\rangle = \sum_{m_{j}\lambda} c_{m_{j}\lambda}\left|m_{j},\lambda\right\rangle
\label{eq:LR-vect-pr-C5}
\end{equation}
where $m_{j}+\lambda$ is a constant, and $c_{m_{j}\lambda}$ are real.
The $C_{6}$ or van der Waals coefficient comes from dipole-dipole interaction whose expression $\hat{V}_{AB}^{dd}(R)$ is obtained by setting $L_{A}=L_{B}=1$ in Eq. (\ref{eq:LR-Potential})
\begin{equation}
\hat{V}_{AB}^{dd}(R)=-\frac{2}{R^{3}}\sum_{M=-1}^{1}\frac{\hat{Q}_{1}^{M}(\hat{r}_{A})\hat{Q}_{1}^{-M}(\hat{r}_{B})}{\left(1+M\right)!\left(1-M\right)!}.\label{eq:LR-pot-dd}
\end{equation}
As the permanent dipole of both the atom and the dimer are zero for any states of the atom and the molecule, the (non-degenerate) second-order perturbation theory is used to obtain the related correction written as $C_{6}/R^{6}$ with
\begin{eqnarray}
C_{6} & = & -4\sum_{a,b\neq0}\frac{1}{\left(E_{a}^{0}-E_{A0}^{0}\right)+\left(E_{b}^{0}-E_{B0}^{0}\right)}\nonumber \\
 & \times & \sum_{M=-1}^{1}\frac{\left\langle \Phi_{A0}^{0}\left|\hat{Q}_{1}^{M}\right|\Phi_{a}^{0}\right\rangle \left\langle \Phi_{B0}^{0}\left|\hat{Q}_{1}^{-M}\right|\Phi_{b}^{0}\right\rangle }{\left(1+M\right)!\left(1-M\right)!}\nonumber \\
 & \times & \sum_{M'=-1}^{1}\frac{\left\langle \Phi_{a}^{0}\left|\hat{Q}_{1}^{-M'}\right|\Phi_{A0}^{0}\right\rangle \left\langle \Phi_{b}^{0}\left|\hat{Q}_{1}^{M'}\right|\Phi_{B0}^{0}\right\rangle }{\left(1+M'\right)!\left(1-M'\right)!}.
\label{eq:LR-C6-gene}
\end{eqnarray}
For a given $\left|\Phi_{0}^{0}\right\rangle $, the corresponding $C_{6}$ coefficient can be expanded as a linear combination of \textit{crossed} terms
\begin{equation}
C_{6}=\sum_{m_{j_1}\lambda_{1}}\sum_{m_{j_2}\lambda_{2}}
c_{m_{j_1}\lambda_{1}}^{}c_{m_{j_2}\lambda_{2}}C_{6}^{cr}\,,
\end{equation}
where the crossed coefficients $C_{6}^{cr}$ are
\begin{eqnarray}
C_{6}^{cr} & = & -4\sum_{a,b}\sum_{M,M'}\frac{1}{\left(E_{a}^{0}-E_{Xv_{d}j}^{0}\right)+\left(E_{b}^{0}-E_{n\ell}^{0}\right)}\nonumber \\
 & \times & \frac{\left\langle Xv_{d}jm_{j_1}\left|\hat{Q}_{1}^{M}\right|\Phi_{a}^{0}\right\rangle \left\langle n\ell\lambda_{1}\left|\hat{Q}_{1}^{-M}\right|\Phi_{b}^{0}\right\rangle }{\left(1+M\right)!\left(1-M\right)!}\nonumber \\
 & \times & \frac{\left\langle \Phi_{a}^{0}\left|\hat{Q}_{1}^{-M'}\right|Xv_{d}jm_{j_2}\right\rangle \left\langle \Phi_{b}^{0}\left|\hat{Q}_{1}^{M'}\right|n\ell\lambda_{2}\right\rangle }{\left(1+M'\right)!\left(1-M'\right)!}\,.
\label{eq:LR-C6-cross}
\end{eqnarray}
The summation is performed over all possible states $\Phi_{a}^{0}$ and $\Phi_{b}^{0}$ of the $A$ and $B$ systems.

With the approach of imaginary frequencies dipole polarizabilities (see e.g. Ref. \cite{spelsberg1993}), the sum in Eq. (\ref{eq:LR-C6-cross}) can be factorized into separated contributions from A and B. For this purpose, we use the identity
\begin{equation}
\frac{1}{\left|x\right|+\left|y\right|}=\frac{2}{\pi}\int_{0}^{+\infty}d\omega\frac{\left|x\right|\left|y\right|}{\left(x^{2}+\omega^{2}\right)\left(y^{2}+\omega^{2}\right)},\label{eq:LR-facto-residus}\,,
\end{equation}
to transform the first term of Eq. (\ref{eq:LR-C6-cross}), with $\left|x\right|=E_{a}^{0}-E_{Xv_{d}j}^{0}$ and $\left|y\right|=E_{b}^{0}-E_{n\ell}^{0}$. This approach is applicable if $E_{a}^{0}-E_{Xv_{d}j}^{0}>0$ and $E_{b}^{0}-E_{n\ell}^{0}>0$. This is always the case for the dimer if it is in its ground rovibronic state. Moreover, if the dimer is homonuclear, radiative transitions between rovibrational levels belonging to the same electronic state are forbidden, so that the identity still holds for any rovibrational level of the $X^1\Sigma_{g}^{+}$ state. In contrast, atom B is in its first electronically excited state. Therefore, taking cesium as an example, Eq. (\ref{eq:LR-facto-residus}) is correct for all transitions, except $6^2P\to6^2S$, for which $E_{b}^{0}-E_{n\ell}^{0}<0$. In the latter case, a similar factorization can be performed, by setting $\left|y\right|=-E_{b}^{0}+E_{n\ell}^{0}$, and by using the following identity
\begin{eqnarray}
\frac{1}{\left|x\right|-\left|y\right|} & = & -\frac{1}{\left|x\right|+\left|y\right|}+\frac{2\left|x\right|}{x^{2}-y^{2}}\nonumber \\
 & = & \frac{2}{\pi}\int_{0}^{+\infty}d\omega\frac{xy}{\left(x^{2}+\omega^{2}\right)\left(y^{2}+\omega^{2}\right)}\nonumber \\
 & + & \frac{2x}{x^{2}-y^{2}}.\label{eq:LR-facto-res-2}
\end{eqnarray}
Inserting Eqs. (\ref{eq:LR-facto-residus}) and (\ref{eq:LR-facto-res-2}) into Eq. (\ref{eq:LR-C6-cross}), and following Ref. \cite{zhu2004}, we obtain
\begin{widetext}
 \begin{eqnarray}
C_{6}^{cr} & = & -\sum_{M=-1}^{1}\sum_{M'=-1}^{1}\frac{4}{\left(1+M\right)!\left(1-M\right)!\left(1+M'\right)!\left(1-M'\right)!}\nonumber \\
 & \times & \left[\frac{1}{2\pi}\int_{0}^{+\infty}d\omega\alpha_{MM'}^{m_{j_1}m_{j_2}}(i\omega)\alpha_{-M-M'}^{\lambda_{1}\lambda_{2}}(i\omega)\right.\nonumber \\
 &  & \left.+\sum_{b}\Theta(-\Delta E_{b}^{0})\alpha_{MM'}^{m_{j_1}m_{j_2}}(\omega=\Delta E_{b}^{0})\right.\nonumber \\
 &  & \left.\times\left\langle n\ell\lambda_{1}\left|\hat{Q}_{1}^{-M}\right|\Phi_{b}^{0}\right\rangle \left\langle \Phi_{b}^{0}\left|\hat{Q}_{1}^{M'}\right|n\ell\lambda_{2}\right\rangle \right],
 \label{eq:LR-C6-plrz}
\end{eqnarray}
where $\Delta E_{b}^{0}=E_{b}^{0}-E_{n\ell}^{0}$, and $\Theta(x)$ is the Heaviside function, equal to 1 for each downwards transition and 0 otherwise. The expression of dipole polarizabilities is generalized to arbitrary frequencies (real or imaginary) according to
\begin{eqnarray}
\alpha_{MM'}^{m_{j_1}m_{j_2}}(z) & = & 2\left(-1\right)^{M}\sum_{a}\frac{\left(E_{a}^{0}-E_{Xv_{d}j}^{0}\right)}{\left(E_{a}^{0}-E_{Xv_{d}j}^{0}\right)^{2}-z^{2}}\nonumber \\
 &  & \times\left\langle Xv_{d}jm_{j_1}\left|\hat{Q}_{1}^{M}\right|\Phi_{a}^{0}\right\rangle \left\langle \Phi_{a}^{0}\left|\hat{Q}_{1}^{-M'}\right|Xv_{d}jm_{j_2}\right\rangle
 \label{eq:LR-alpha-A-1}\\
\alpha_{-M-M'}^{\lambda_{1}\lambda_{2}}(z) & = & 2\left(-1\right)^{M}\sum_{b}\frac{\left(E_{b}^{0}-E_{n\ell}^{0}\right)}{\left(E_{b}^{0}-E_{n\ell}^{0}\right)^{2}-z^{2}}\nonumber \\
 &  & \times\left\langle n\ell\lambda_{1}\left|\hat{Q}_{1}^{-M}\right|\Phi_{b}^{0}\right\rangle \left\langle \Phi_{b}^{0}\left|\hat{Q}_{1}^{M'}\right|n\ell\lambda_{2}\right\rangle .
 \label{eq:LR-alpha-B-1}
 \end{eqnarray}
\end{widetext}
where $z$ can be real or imaginary. In Eq.(\ref{eq:LR-C6-plrz}) the first term inside the brackets is the well-known product of the polarizabilities at imaginary frequencies of the molecular $|Xv_{d}j \rangle$ and atomic $|n\ell \rangle$ states (the corresponding labels are omitted for clarity sake). In the second term appears  the polarizability of the dimer at the real frequency of each downwards transition of the atom. We note that if the dimer polarizability is negative and significantly large,  the second term in the square brackets can become larger in magnitude than the first term and, therefore can turn the $C_{6}$ sign to positive.

\section{Calculation of polarizabilities}
\label{sec:polar}
\subsubsection{The polarizability of the dimer}
Molecular polarizabilities are most often calculated ignoring rotation, so that it is wise to separate it in the equations of Section \ref{sec:c6}. We start by expressing the dipole matrix element $\left\langle X\Lambda v_{d}jm_{j}\left|\hat{Q}_{1}^{M}\right|\Phi_{a}^{0}\right\rangle $ of Eq. (\ref{eq:LR-alpha-A-1}) between the states defined with respect to the T-CS, in terms of matrix elements between states defined in the D-CS. For clarity, we explicitly write the dimer quantum number $\Lambda$, which is the projection of the electronic angular momentum on the $Z_A$ axis. Specifying as well all the relevant quantum numbers according to $\left|\Phi_{a}^{0}\right\rangle =\left|X'\Lambda'v_{d}'j'm'_{j}\right\rangle $, we have
\begin{eqnarray}
 &  & \left\langle X\Lambda v_{d}jm_{j}\left|\hat{Q}_{1}^{M}\right|\Phi_{a}^{0}\right\rangle \nonumber \\
 & = & \sum_{\mu}\left\langle X\Lambda v_{d}jm_{j}\left|d_{M\mu}^{1}(\delta)\hat{q}_{1}^{\mu}\right|X'\Lambda'v_{d}'j'm'_{j}\right\rangle \nonumber \\
 & = & \sum_{\mu}\left\langle jm_{j}\left|d_{M\mu}^{1}\right|j'm'_{j}\right\rangle \left\langle X\Lambda v_{d}\left|\hat{q}_{1}^{\mu}\right|X'\Lambda'v_{d}'\right\rangle ,\label{eq:LR-C6-dip-A}
\end{eqnarray}
where the quantities $\hat{q}_{1}^{\mu}$ are the components of the electric dipole moment defined with respect to the D-CS. The index $\mu$ is either 0 for $\Sigma\to\Sigma$ transitions ($\Lambda=\Lambda'=0$), or $\pm 1$ for $\Sigma\to\Pi$ transitions. In the T-CS the rotational wave function of the dimer is proportional to $d_{m_{j}0}^{j}$ and to $d_{m_{j}\pm1}^{j}$, for $\Sigma$ and $\Pi$ states, respectively. Using the properties of the $d^j_{\mu \nu}$ rotation matrices \cite{varshalovich1988}, we find
\begin{equation}
\left\langle jm_{j}\left|d_{M\mu}^{1}\right|j'm'_{j}\right\rangle =\sqrt{\frac{2j'+1}{2j+1}}C_{1Mj'm'_{j}}^{jm_{j}}C_{1\mu j'-\mu}^{j0}.
\label{eq:LR-C6-dip-A-rot}
\end{equation}
where the $C_{1Mj'm'_{j}}^{jm_{j}}$ and $C_{1\mu j'-\mu}^{j0}$ are Clebsch-Gordan coefficients. If in Eq. (\ref{eq:LR-alpha-A-1}) we neglect the rotational part of the $0^{\rm th}$-order energy (only low rotational levels are relevant for the cold temperature domain), i.e.
\begin{equation}
E_{Xv_{d}j}^{0}=E_{Xv_{d}}^{0}+B_{v_{d}}j\left(j+1\right)\approx E_{Xv_{d}}^{0},
\end{equation}
we identify the exact expressions of the parallel polarizability $\alpha_{\parallel}$ and of the perpendicular polarizability $\alpha_{\bot}$ with respect to the $Z_A$ axis, for $\mu=0$ and $\pm1$ respectively. Therefore, Eq. (\ref{eq:LR-alpha-A-1}) becomes
\begin{eqnarray}
\alpha_{MM'}^{m_{j_1}m_{j_2}}(z) & \approx & \sum_{j',m'_{j}}\frac{2j+1}{2j'+1} \left[\left(C_{10j0}^{j'0}\right)^{2}\alpha_{\parallel}(z)\right. \left.+2\left(C_{11j0}^{j'1}\right)^{2}\alpha_{\bot}(z)\right]\nonumber \\
 & \times & C_{1-Mjm_{j_1}}^{j'm'_{j}}C_{1-M'jm_{j_2}}^{j'm'_{j}},
 \label{eq:LR-alpha-A-2}
\end{eqnarray}
where we used the identity \cite{varshalovich1988}
\begin{equation}
C_{a\alpha b\beta}^{c\gamma}=\left(-1\right)^{a-\alpha}\sqrt{\frac{2c+1}{2b+1}}C_{c\gamma a-\alpha}^{b\beta}
\label{eq:LR-CG-invers-ind}\,,
\end{equation}
in order to put all the primes in the upper indices.
The dependence on the rotational state of the dimer is restricted to the coefficients of the vibronic polarizabilities of the dimer. We note that Eq. (\ref{eq:LR-alpha-A-2}) is valid for real, imaginary, and zero frequencies (\textit{i.e.} for static polarizabilities).
\subsubsection{The valence contribution to the atomic polarizability}
We perform a similar development for the atom, in order to separate the radial and angular parts of the matrix elements of the dipole moment. By writing explicitly $\left|\Phi_{b}^{0}\right\rangle =\left|n'\ell'\lambda'\right\rangle $ and by using Eq. (\ref{eq:LR-CG-invers-ind}), we rewrite the dipole matrix elements as
\begin{equation}
\left\langle n\ell\lambda\left|\hat{Q}_{1}^{-M}\right|n'\ell'\lambda'\right\rangle =\left(-1\right)^{M}\sqrt{\frac{2\ell+1}{2\ell'+1}}C_{10\ell0}^{\ell'0}C_{1M\ell\lambda}^{\ell'\lambda'}r_{n\ell n'\ell'}
\end{equation}
where $r_{n\ell n'\ell'}=\left\langle n\ell\left|r\right|n'\ell'\right\rangle $ is the matrix element of the valence electron radial coordinate. Eq. (\ref{eq:LR-alpha-B-1}) becomes
\begin{eqnarray}
\alpha_{-M-M'}^{\lambda_{1}\lambda_{2}}(z) & = & 2\sum_{n',\ell'}\frac{\left(E_{n'\ell'}^{0}-E_{n\ell}^{0}\right)}{\left(E_{n'\ell'}^{0}-E_{n\ell}^{0}\right)^{2}-z^{2}}\nonumber \\
 &  & \times r_{n\ell n'\ell'}^{2}\frac{2\ell+1}{2\ell'+1}\left(C_{10\ell0}^{\ell'0}\right)^{2}\nonumber \\
 &  & \times\sum_{\lambda'}C_{1M\ell\lambda_{1}}^{\ell'\lambda'}C_{1M'\ell\lambda_{2}}^{\ell'\lambda'},
\label{eq:LR-alpha-B-2}
\end{eqnarray}
The index $\ell'$ above takes values corresponding to dipole-allowed transitions, \textit{i.e.} $\ell'=0, 2$ in the present case of a Cs($6^2P$) atom, and $n'$ for all the relevant atomic levels.

The similarity between Eqs. (\ref{eq:LR-alpha-A-2}) and (\ref{eq:LR-alpha-B-2}) confirms the equivalence in the formalism between the rotational angular momentum of the dimer and the electronic orbital momentum of the atom, which makes the generalization to more complicated cases like molecule-molecule long-range interactions quite straightforward. However, it is not possible to express Eq. (\ref{eq:LR-alpha-B-2}) as a function of the sole polarizability of the atomic $n\ell$ level. For instance, if $\ell=1$, $M'=M$ and $\lambda_{2}=\lambda_{1}$, the dipole polarizability $\alpha_{MM}^{\lambda_{1}\lambda_{1}}(z)$ of the sublevel $nP\lambda_{1}$ contains angular factors which are different for $P\to S$ and $P\to D$ transitions \cite{zhu2004}. In the usual definition of the polarizability, an average is made over all the sublevels $\lambda_{1}$ leading to the disappearance of  the angular factors which is not the case here. Therefore, we introduce \textit{state-to-state} polarizabilities $\alpha_{n\ell n'\ell'}$ associated to the different $n\ell\to n'\ell'$ transitions
\begin{eqnarray}
\alpha_{n\ell n'\ell'}(z) & = & \frac{2}{3}\frac{\left(E_{n'\ell'}^{0}-E_{n\ell}^{0}\right)}{\left(E_{n'\ell'}^{0}-E_{n\ell}^{0}\right)^{2}-z^{2}}\nonumber \\
 &  & \times r_{n\ell n'\ell'}^{2}\left(C_{10\ell0}^{\ell'0}\right)^{2},
\label{eq:LR-alpha-ell-ellp}
\end{eqnarray}
so that Eq. (\ref{eq:LR-alpha-B-2}) becomes
\begin{eqnarray}
\alpha_{-M-M'}^{\lambda_{1}\lambda_{2}}(z) & = & 3 \sum_{\ell'=\{\ell-1,\ell+1\}} \frac{2\ell+1}{2\ell'+1} \sum_{n'} \alpha_{n\ell n'\ell'}(z)\nonumber \\
 & \times & \sum_{\lambda'}C_{1M\ell\lambda_{1}}^{\ell'\lambda'}C_{1M'\ell\lambda_{2}}^{\ell'\lambda'} \,.
\end{eqnarray}
The state-to-state polarizabilities obey the property $\sum_{n'\ell'}\alpha_{n\ell n'\ell'}\approx\alpha_{n\ell}$, with $\alpha_{n\ell}$ being the actual (isotropic) atomic polarizability of the level $n\ell$. This identity is only approximate as the effect of core electrons have been neglected so far.
\subsubsection{The core contribution to the atomic polarizability}
Following Ref. \cite{derevianko2010} we assume that the contribution of the core electrons can be treated as an additional correction $\alpha_{c}$ to the total polarizability $\bar{\alpha}_{n\ell}$, independent of the atomic state $|n\ell \rangle$
\begin{equation}
\bar{\alpha}_{n\ell}=\sum_{n'\ell'}\alpha_{n\ell n'\ell'}+\alpha_{c} \,.
\end{equation}
In our numerical calculations, $\alpha_{c}$ is obtained in the following way. First, the contribution of the valence electron is evaluated using tabulated values of dipole moments for transitions from the $6^2S$ state of cesium. The result is then compared to the polarizability of Ref. \cite{derevianko2010} which accounts for the ionic core contribution, and the difference between our result and that of Ref. \cite{derevianko2010} is then attributed to  $\alpha_{c}$.

It is not straightforward to see that the core polarizability brings a simple additive contribution to to the $C_6$ coefficients. As inner shells have different angular quantum numbers,  we first consider only the electrons of a given closed inner shell, \textit{e.g.} $4d$ for cesium. A sum over all the closed shells will be taken at the end. We note $n_{c}$, $\ell_{c}$, $k_{i}$, $\sigma_{i}$ the principal, orbital, azimuthal and spin quantum numbers of the $i^{\textrm{th}}$ electron of the shell. The indexes $n_{c}$ and $\ell_{c}$ are identical for the $2\left(2\ell_{c}+1\right)$ electrons, whereas $k_{i}$ varies from $-\ell_{c}$ to $+\ell_{c}$, and $\sigma_{i}=\pm\frac{1}{2}$. Since the dipole operator is mono-electronic, the ionic core brings to the $C_{6}$ coefficient an additional term $C_{6}^{c}$, which is the sum of all individual contributions
\begin{equation}
C_{6}^{n_cl_c}=\sum_{i\in\textrm{shell}}C_{6}^{n_cl_c}(i)\,,
\end{equation}
where the general expression for $C_{6}^{n_cl_c}(i)$ is
\begin{eqnarray}
C_{6}^{n_cl_c}(i) & = & -4\sum_{a,b\neq0}\sum_{M,M'}\frac{1}{\left(E_{a}^{0}-E_{A0}^{0}\right)+\left(E_{b}^{0}-E_{B0}^{0}\right)}\nonumber \\
 & \times & \frac{\left\langle \Phi_{A0}^{0}\left|\hat{Q}_{1}^{M}\right|\Phi_{a}^{0}\right\rangle \left\langle \Phi_{B0}^{0}\left|\hat{Q}_{1}^{-M}(i)\right|\Phi_{bi}^{0}\right\rangle }{\left(1+M\right)!\left(1-M\right)!}\nonumber \\
 & \times & \frac{\left\langle \Phi_{a}^{0}\left|\hat{Q}_{1}^{-M'}\right|\Phi_{A0}^{0}\right\rangle \left\langle \Phi_{bi}^{0}\left|\hat{Q}_{1}^{M'}(i)\right|\Phi_{B0}^{0}\right\rangle }{\left(1+M'\right)!\left(1-M'\right)!},
 \label{eq:LR-C6-core}
\end{eqnarray}
In Eq. (\ref{eq:LR-C6-core}), the atomic states are characterized by independent electrons
\begin{eqnarray}
\left|\Phi_{B0}^{0}\right\rangle  & = & \left|n_{c}\ell_{c},-\ell_{c},-\frac{1}{2}\right\rangle \times\dots\nonumber \\
 & \times & \left|n_{c}\ell_{c},k_{i},\sigma_{i}\right\rangle \times\dots\nonumber \\
 & \times & \left|n_{c}\ell_{c},\ell_{c},\frac{1}{2}\right\rangle \left|n\ell\lambda\right\rangle\,,
\end{eqnarray}
and similarly $\left|\Phi_{bi}^{0}\right\rangle =\left|n_{c}\ell_{c},-\ell_{c},-\frac{1}{2}\right\rangle \dots\left|n'_{c}\ell'_{c},k'_{i},\sigma_{i}\right\rangle \dots\left|n'_{c}\ell'_{c},\ell'_{c},\frac{1}{2}\right\rangle \left|n\ell\lambda\right\rangle $, and the molecular states, $\left|\Phi_{A0}^{0}\right\rangle $ and $\left|\Phi_{a}^{0}\right\rangle $, have the same quantum numbers as previously. The states $\left|\Phi_{B0}^{0}\right\rangle $ and $\left|\Phi_{bi}^{0}\right\rangle$ are antisymmetric with respect to the permutation of two electrons. However, for simplicity, they are written here as simple products of the individual electronic states, which does not modify the value of $C_6^{n_cl_c}$. In Eq.(\ref{eq:LR-C6-core}), the energy $E_{b}^{0}$ is independent on the index $i$, since $i$ only makes a distinction between different projection of orbital and spin angular momenta.

Unlike the valence electron, as the core is not affected by the $C_{5}$ coefficient, there is no crossed term in the calculation of $C_{6}^{n_cl_c}(i)$. As a consequence, the Wigner-Eckart theorem imposes $m'_{j}+M=m_{j}$ and $m'_{j}+M'=m_{j}$, as well as $k_{i}'-M=k_{i}$ and $k_{i}'+M'=k_{i}$, hence the condition $M'=M$. So, the double summation of Eq. (\ref{eq:LR-C6-core}) reduces to a single one over $M$. Similarly to Eq. (\ref{eq:LR-C6-plrz}), we can factorize Eq. (\ref{eq:LR-C6-core}) with polarizability-like quantities. Because there is no allowed transitions from the last inner shell of the atom to lower-energy states, that factorization does not contain any additional term due to the excited state. It yields
\begin{eqnarray}
C_{6}^{n_cl_c} & = & -\frac{2}{\pi}\sum_{M=-1}^{1}\frac{1}{\left[\left(1+M\right)!\left(1-M\right)!\right]^{2}}\nonumber \\
 & \times & \int_{0}^{+\infty}d\omega\alpha_{MM}^{m_{j}m_{j}}(i\omega)\alpha_{-M-M}^{n_cl_c}(i\omega)\,,
\end{eqnarray}
where $\alpha_{MM}^{m_{j}m_{j}}$ is given by Eq. (\ref{eq:LR-alpha-A-2}) and $\alpha_{-M-M}^{n_cl_c}$ is similar to (\ref{eq:LR-alpha-B-2}):
\begin{eqnarray}
\alpha_{-M-M}^{n_cl_c}(i\omega) & = & 2\sum_{n'_{c}\ell'_{c}}
\frac{E_{n'_{c}\ell'_{c}}^{0}-E_{n_{c}\ell_{c}}^{0}} {\left(E_{n'_{c}\ell'_{c}}^{0}-E_{n_{c}\ell_{c}}^{0}\right)^2+\omega^2} \nonumber \\
 & \times & \frac{2\ell_{c}+1}{2\ell'_{c}+1}\left(C_{10\ell_{c}0}^{\ell'_{c}0}\right)^{2}\left(r_{n_{c}\ell_{c}n'_{c}\ell'_{c}}\right)^{2}\nonumber \\
 & \times & \sum_{kk'\sigma\sigma'}\left(C_{1M\ell_{c}k}^{\ell'_{c}k'}\right)^{2}\delta_{\sigma\sigma'}.
 \label{eq:LR-alpha-Bc-1}
\end{eqnarray}
In Eq. (\ref{eq:LR-alpha-Bc-1}), the sum over $k$ and $\sigma$ runs over all the core electrons $i$. The sum over the two different spin projections for each orbital gives factor of 2. The sum of the Clebsch-Gordan coefficients yields $\frac{2\ell'_{c}+1}{3}$. Therefore, the angular part of the dipole moment disappears and $\alpha_{-M-M}^{c}$ finally reads
\begin{eqnarray}
\alpha_{-M-M}^{n_cl_c}(i\omega) & = & 4\sum_{n'_{c}\ell'_{c}} \frac{E_{n'_{c}\ell'_{c}}^{0}-E_{n_{c}\ell_{c}}^{0}} {\left(E_{n'_{c}\ell'_{c}}^{0}-E_{n_{c}\ell_{c}}^{0}\right)^2+\omega^2} \nonumber \\
  & \times & \frac{2\ell_{c}+1}{3}
\left(C_{10\ell_{c}0}^{\ell'_{c}0}\right)^{2} \left(r_{n_{c}\ell_{c}n'_{c}\ell'_{c}}\right)^{2}\,.
\end{eqnarray}
As one can see from the above expression, it is not necessary to separate  different series of transitions $\ell_{c}\to\ell_{c}'$, and we obtain a meaningful polarizability. Finally, we can extend this results to all inner shells that gives the core polarizability
\begin{eqnarray}
\alpha_{c}(i\omega) & = & 4\sum_{n_{c}\ell_{c}n'_{c}\ell'_{c}} \frac{E_{n'_{c}\ell'_{c}}^{0}-E_{n_{c}\ell_{c}}^{0}} {\left(E_{n'_{c}\ell'_{c}}^{0}-E_{n_{c}\ell_{c}}^{0}\right)^2+\omega^2} \nonumber \\
  & \times & \frac{2\ell_{c}+1}{3}
\left(C_{10\ell_{c}0}^{\ell'_{c}0}\right)^{2} \left(r_{n_{c}\ell_{c}n'_{c}\ell'_{c}}\right)^{2}\,.
\end{eqnarray}
and the core contribution $C_{6}^{c}$ to the $C_{6}$ coefficient
\begin{eqnarray}
\label{eq:LR-C6-core-2}
C_{6}^{c} & = & -\frac{2}{\pi}\sum_{M}\frac{1}{\left[\left(1+M\right)!\left(1-M\right)!\right]^{2}}\nonumber \\
 & \times & \sum_{j'm_{j'}}\frac{2j+1}{2j'+1}\left(C_{1Mjm_{j}}^{j'm'_{j}}\right)^{2}\int d\omega\alpha_{c}(i\omega)\nonumber \\
 &  & \times\left(\left(C_{10j0}^{j'0}\right)^{2}\alpha_{\parallel}(i\omega)+2\left(C_{11j0}^{j'1}\right)^{2}\alpha_{\bot}(i\omega)\right)\,.
\end{eqnarray}
With the approximations above, we obtain an additional term which depends on the physical polarizabilities of the atomic core and the molecule, and in which the rotation of the dimer appears explicitly. As all the factors in Eq. (\ref{eq:LR-C6-core-2}) are positive, $C_{6}^{c}$ is negative and, thus, it makes the interaction between the atom and the dimer more attractive (or less repulsive).
\subsubsection{General expression for $C_{6}$}
We summarize all the results of the previous paragraphs. We recall that the $C_{6}$ coefficients are calculated for each eigenvector of the quadrupole-quadrupole operator. Each eigenvector $\left|\Phi_{0}^{0}\right\rangle $ is a linear combination of states $\left|m_{j},\lambda\right\rangle $ (see Eq. (\ref{eq:LR-vect-pr-C5})) with $m_{j}+\lambda$ being constant. For a given $\left|\Phi_{0}^{0}\right\rangle $, the general expression for $C_{6}$ is
\begin{widetext}\begin{eqnarray}
C_{6} & = & -3\sum_{m_{j_1}\lambda_{1}}\sum_{m_{j_2}\lambda_{2}}\sum_{MM'}\sum_{j'm'_{j}}\sum_{\ell'\lambda'}\frac{c_{m_{j_1}\lambda_1}c_{m_{j_2}\lambda_2}}{\left(1+M\right)!\left(1-M\right)!\left(1+M'\right)!\left(1-M'\right)!}\nonumber \\
 & \times & \frac{2j+1}{2j'+1}C_{1-Mjm_{j_1}}^{j'm'_{j}}C_{1-M'jm_{j_2}}^{j'm'_{j}}\frac{2\ell+1}{2\ell'+1} C_{1M\ell\lambda_{1}}^{\ell'\lambda'}C_{1M'\ell\lambda_{2}}^{\ell'\lambda'}\nonumber \\
 & \times & \left[\frac{2}{\pi}\int_{0}^{+\infty}d\omega\left(\left(C_{10j0}^{j'0}\right)^{2}\alpha_{\parallel}(i\omega)+2\left(C_{11j0}^{j'1}\right)^{2}\alpha_{\bot}(i\omega)\right)\alpha_{\ell\ell'}(i\omega)\right.\nonumber \\
 &  & \left.+4\sum_{n'}\Theta(-\Delta E_{n'\ell',n\ell}^{0})
\left(\left(C_{10j0}^{j'0}\right)^{2}\alpha_{\parallel}(\Delta E_{n'\ell',n\ell}^{0}) +2\left(C_{11j0}^{j'1}\right)^{2}\alpha_{\bot}(\Delta E_{n'\ell',n\ell}^{0}) \right)\left(\mu_{n'\ell',n\ell}\right)^{2}\right]\nonumber \\
 & - & \frac{2}{\pi}\sum_{m_{j_1}\lambda_{1}} \sum_{M}\sum_{j'm_{j}'}\frac{c_{m_{j_1}\lambda_{1}}^{2}}{\left[\left(1+M\right)!\left(1-M\right)!\right]^{2}}\frac{2j+1}{2j'+1}\left(C_{1Mjm_{j1}}^{j'm'_{j}}\right)^{2}\nonumber \\
 & \times & \int_{0}^{+\infty}d\omega\left(\left(C_{10j0}^{j'0}\right)^{2}\alpha_{\parallel}(i\omega)+2\left(C_{11j0}^{j'1}\right)^{2}\alpha_{\bot}(i\omega)\right)\alpha_{c}(i\omega),
\label{eq:LR-C6-total}
\end{eqnarray}
\end{widetext}
where
\begin{equation}
\mu_{n'\ell',n\ell}=\frac{1}{\sqrt{3}}r_{n'\ell',n\ell}C_{10\ell0}^{\ell'0}
\end{equation}
is the atomic transition dipole moment, $\Theta(x)$ is Heaviside's function, and $\Delta E_{n'\ell',n\ell}^{0}=E_{n'\ell'}^{0}-E_{n\ell}^{0}$.

\section{Long-range potential curves for the Cs-Cs$_2$ complex}
\label{sec:results}

The atomic polarizability of Cs($6^2P$) comes from the accurate calculations of transition dipole moments in Ref.\cite{iskrenova-tchoukova2007}, averaged over the $6P_{1/2}$ and $6P_{3/2}$ levels. The atomic transition energies are extracted from Ref.\cite{moore1952}, and averaged over fine structure levels. In Eq.(\ref{eq:LR-alpha-ell-ellp}), the summation is restricted to $n'=7$ to 10 for $\ell'=0$ and $n'=5$ to 10 for $\ell'=2$. We use a mixture of experimental and theoretical molecular data to compute the molecular polarizabilities at imaginary frequencies, which involve summation over all vibrational levels of all electronic states of Cs$_2$. The Cs$_2$ ground state potential curve is taken from Ref.\cite{amiot2002}, and the one of the $B^1\Pi_u(6s+6p)$ state from Ref.\cite{diemer1989}. The $A^1\Sigma_u^+(6s+6p)$ and $b^3\Pi_u(6s+6p)$ and their spin-orbit coupling are those fitted to reproduce the data of Ref.\cite{danzl2008}. All the other electronic states come from quantum chemistry calculations performed in our group according to the method described in Ref.\cite{aymar2005} as well as transition dipole moments which will be presented in a separate publication. The dissociation energies of the atom-dimer complex are given by Cs$_2$ rotational energies, $B_{0}j\left(j+1\right)$ where the rotational constant for the $v_d=0$ level of Cs$_{2}$ is $B_{0}=1.17314\times10^{-2}$ cm$^{-1}$ \cite{amiot2002}.

\begin{figure}
\begin{centering}
\includegraphics[width=0.6\textwidth]{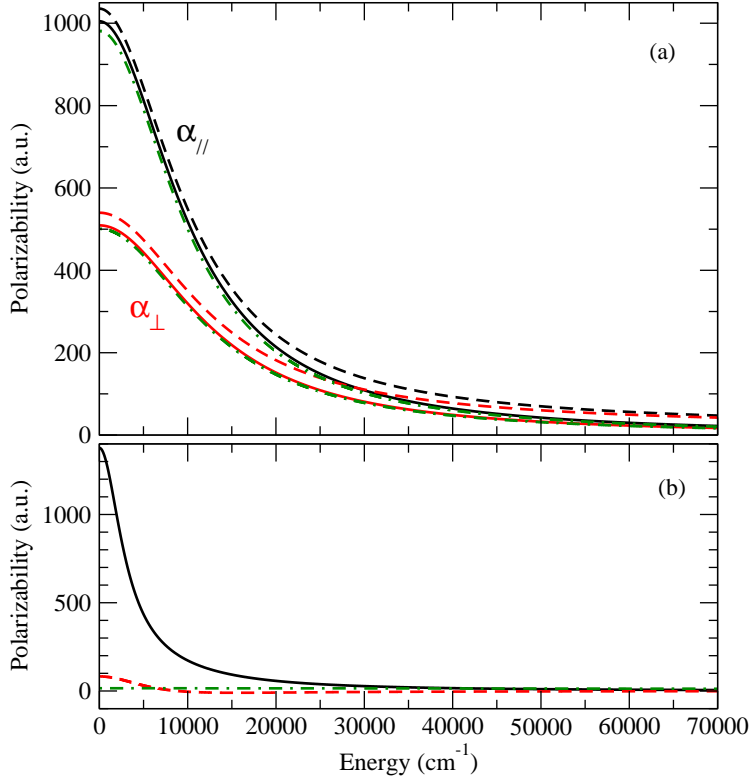}
\par\end{centering}
\caption{Calculated dipole polarizabilities in imaginary frequencies (a) for Cs$_2(X^1\Sigma_g^+, v_d=0)$ (with the parallel and perpendicular components, full lines), and (b) for Cs$(6^2P)$ (lower panel, full line). The contribution of the sole $B^1\Pi_u(6s+6p)$ state to $\alpha_{\bot}$, and of the pair of $0_u^+$ $A^1\Sigma_u^+(6s+6p)$ and $b^3\Pi_u(6s+6p)$ states to $\alpha_{\parallel}$, are shown (dot-dashed lines). (a): The polarizability components including the core polarizability are drawn with dashed lines. (b) The magnitude of second term of Eq.\ref{eq:LR-C6-plrz} (dashed line), and of the core polarizability (dot-dashed line) are also shown.}
\label{fig:polar_im}
\end{figure}

The variation of the dipole polarizability at imaginary frequencies is displayed in Fig. \ref{fig:polar_im}. The $B^1\Pi_u(6s+6p)$ state lowest electronic state contributes to the valence part for about 99\% to $\alpha_{\bot}$, as well as the pair of states $A^1\Sigma_u^+(6s+6p)$ and $b^3\Pi_u(6s+6p)$ coupled by spin-orbit interaction to yield a pair of so-called $0_u^+$ states. The core polarizability represents only a few percents of the total one, for the represented frequency domain. We numerically checked that the dissociation continua can be neglected.

The $C_{6}$ coefficients are first calculated for Cesium in its ground state $6^2S$ interacting with Cs$_2$ $X^1\Sigma_g^+(v_d=0,j)$ molecule, for which the quadrupole-quadrupole interaction is zero. The $\alpha_{\ell\ell'}$ quantities now represent the actual dipole polarizability. Moreover there is no crossed term in the $C_{6}$ coefficients and Eq. (\ref{eq:LR-C6-total}) reduces to
\begin{eqnarray}
C_{6} & = & -\frac{2}{\pi}\sum_{M}\sum_{j'm_{j}'}\frac{1}{\left[\left(1+M\right)!\left(1-M\right)!\right]^{2}}\nonumber \\
 & \times & \frac{2j+1}{2j'+1}\left(C_{1Mjm_{j}}^{j'm'_{j}}\right)^{2}\int_{0}^{+\infty}d\omega\alpha(i\omega)\nonumber \\
 & \times & \left(\left(C_{10j0}^{j'0}\right)^{2}\alpha_{\parallel}(i\omega)+2\left(C_{11j0}^{j'1}\right)^{2}\alpha_{\bot}(i\omega)\right)\,,
\label{eq:C6-Cs6s}
\end{eqnarray}
where $\alpha(i\omega)$ is the atomic polarizability in the $6^2S$ state, including core contributions. It is worth noting that in the particular case $j=0$, Eq.(\ref{eq:C6-Cs6s}) can be written in a similar form as for two $S$ atoms
\begin{eqnarray}
C_{6} = -\frac{3}{\pi} \int_{0}^{+\infty}d\omega\alpha(i\omega)\bar{\alpha}(i\omega)\,,
\label{eq:C6_SS}
\end{eqnarray}
with $\bar{\alpha}=(\alpha_{\parallel}+2\alpha_{\bot})/3$ the so-called isotropic polarizability of the dimer.

\begin{table}
\begin{centering}
\begin{tabular}{|c|c|c||c|c|c|}
\hline
symmetry & ~~$j$~~ & $C_{6}$ (a.u.) &
symmetry & ~~$j$~~ & $C_{6}$ (a.u.) \tabularnewline
\hline
\hline
$\Sigma^{+}$ &  0  & -12101 &
$\Pi$  & 4 &  -12587  \tabularnewline
          & 1 &  -12981  &
$\Delta$ & 2 &  -11473  \tabularnewline
          & 2 &  -12729  &
          & 3 &  -12101   \tabularnewline
          & 3 &  -12688  &
          &  4  &  -12330  \tabularnewline
          &  4  &  -12672  &
$\Phi$&  3  &   -11369  \tabularnewline
$\Pi$  & 1 & -11662  &
          & 4 &  -11902   \tabularnewline
          & 2 &  -12415  &
$\Gamma$ & 4 &  -11302  \tabularnewline
          & 3 & -12541  &
          &   &    \tabularnewline
\hline
\end{tabular}
\par\end{centering}
\caption{\label{tab:C6-Cs6s} The $C_6$ coefficients of the Cs$_2(X^1\Sigma_g^+, v_d=0, j)$+Cs($6^2S$) long-range interaction calculated
for $j=0$ to 4. In analogy to a diatomic molecule, the $C_6$ are sorted by projections of the total orbital quantum number $m_J=m_j$ on the $Z$ axis (note that $\lambda=0$), and the parity $\pm$ through the reflection symmetry with respect any plane containing this axis.}
\end{table}

\begin{figure}
\begin{centering}
\includegraphics[width=0.6\textwidth]{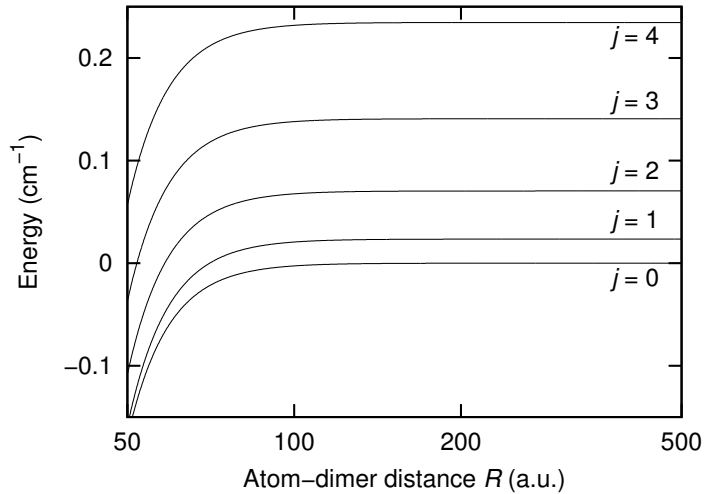}
\par\end{centering}
\caption{Long-range potential energy curves $B_{0}j(j+1) + C_{6}/R^{6}$ of $\Sigma^+$ symmetry, as functions of the atom-dimer distance $R$ (in logarithmic scale), representing the interaction between Cs($6^2S$) and Cs$_2(X^1\Sigma_g^+, v_d=0, j)$ for $j=0$ to 4. The related $C_6$ coefficients are given in Table \ref{tab:C6-Cs6s}.}
\label{fig:C6-Cs6s}
\end{figure}

The results of our calculations are given in Table \ref{tab:C6-Cs6s} and in Figure \ref{fig:C6-Cs6s}. They are sorted by values of the projection $m_J$ of the total orbital momentum on the $Z$ axis, and for $m_J=0$, the parity $\pm$ through the reflection symmetry with respect any plane containing this axis. In analogy to diatomic molecules, these symmetries are labeled $\Sigma^{\pm},\,\Pi,\,\Delta$... All the $C_6$ coefficients are negative, reflecting an attraction between the atom and the dimer which is slightly less than twice the one for two interacting Cs$(6^2S)$ atoms (6840~a.u. \cite{derevianko1999,derevianko2001,derevianko2010}). This can be understood from the simple form of Eq.(\ref{eq:C6_SS}), considering that the static dipole polarizability of a ground state Cs atom and of a ground state Cs$_2$ molecule are respectively $\approx$402~a.u. \cite{derevianko1999}, $\approx$707~a.u. \cite{deiglmayr2008},\footnote{The quantity $2\alpha_c$(Cs$^+$)=30.8~a.u. deduced from Ref.\cite{derevianko1999} has been added to the one of Ref.\cite{deiglmayr2008}}.  In addition, no curve crossing are visible on Figure \ref{fig:C6-Cs6s} as all $C_6$ coefficients have close values. Therefore, the validity of the long-range expansion is limited at short distances by the overlap of the electronic clouds. With $\left\langle r_{6S}^2\right\rangle=42$ a.u. \cite{aymar-pc}, we estimate the value of the LeRoy radius to $R_{LR}\sim40-45$ a.u..

\begin{table}
\begin{centering}
\begin{tabular}{|c|c|c|c||c|c|c|c|}
\hline
symmetry & ~~$j$~~ & $C_{5}$ (a.u.) & $C_{6}$ (a.u.) &
symmetry & ~~$j$~~ & $C_{5}$ (a.u.) & $C_{6}$ (a.u.) \tabularnewline
\hline
\hline
$\Sigma^{+}$ &  0  &  0  &  -42704  &
$\Pi$  & 4 & -739 &  671  \tabularnewline
          & 1 &  -1674  &  51249  &
          & 4 &  108 &  -15884   \tabularnewline
          & 1 &  0  &  -21562  &
          & 4 &  522 &  -47279   \tabularnewline
          &  2  &  -913  &  12128  &
$\Delta $  &  1  &  -279  &  -18694   \tabularnewline
          &  2  &  116  & -16885  &
          &  2  &  -140  &  -21244  \tabularnewline
          & 3 & -796  &   4923  &
          & 2 & 1136  &  -95614  \tabularnewline
          & 3 &  145  &  -15420  &
          & 3 & -835  &  -1624  \tabularnewline
          & 4 & -755  &  2251  &
          & 3 & -87   &  -19563  \tabularnewline
          & 4 &  157  &  -14835  &
          & 3 &  736  &  -65454  \tabularnewline
$\Sigma^{-}$ & 1 &  0  &  -43920  &
          & 4 & -721  &  -2643  \tabularnewline
          & 2 &  399  &  -45131  &
          & 4 & -11  &  -17153  \tabularnewline
          & 3 &  465  &  -45333  &
          & 4 &  623  &  -56200  \tabularnewline
          & 4 &  489  &  -45407  &
$\Phi$   & 2 & -399  &  -16589  \tabularnewline
$\Pi $ &  0  &  0  &  -23605  &
          & 3 & -245  &  -18030  \tabularnewline
          & 1 &  0  &  -29303  &
          & 3 & 1175  &  -103161  \tabularnewline
          & 1  &  1116  &  -79756  &
          & 4 & -783  &  -5057  \tabularnewline
          & 2 & -964  &  7305  &
          & 4 & -161  &  -17444  \tabularnewline
          & 2 & -19 &  -22961  &
          & 4 &  835  &  -76003  \tabularnewline
          & 2 &  584  &  -50736  &
$\Gamma$ & 3 & -465  &  -15420  \tabularnewline
          & 3 & -783  &  2496  &
          & 4 & -320  &  -16392  \tabularnewline
          & 3 &  64 &  -17295  &
          & 4 & 1208  &  -107555  \tabularnewline
          & 3 &  532  &  -48103  &
H        & 4 & -507  &  -14676  \tabularnewline\hline
\end{tabular}
\par\end{centering}
\caption{\label{tab:LR-C5-C6} The $C_{5}$ and $C_6$ coefficients of the Cs$_2(X^1\Sigma_g^+, v_d=0, j)$+Cs($6^2P$) long-range interaction calculated
for $j=0$ to 4. In analogy to a diatomic molecule, the states are sorted by projections of the total orbital quantum number $m_J=m_j+\lambda$ on the $Z$ axis, and the $\pm$ reflection symmetry through any plane containing this axis. The values for $C_5$ are taken from Paper I.}
\end{table}

\begin{figure}
\begin{centering}
\includegraphics[width=0.6\textwidth]{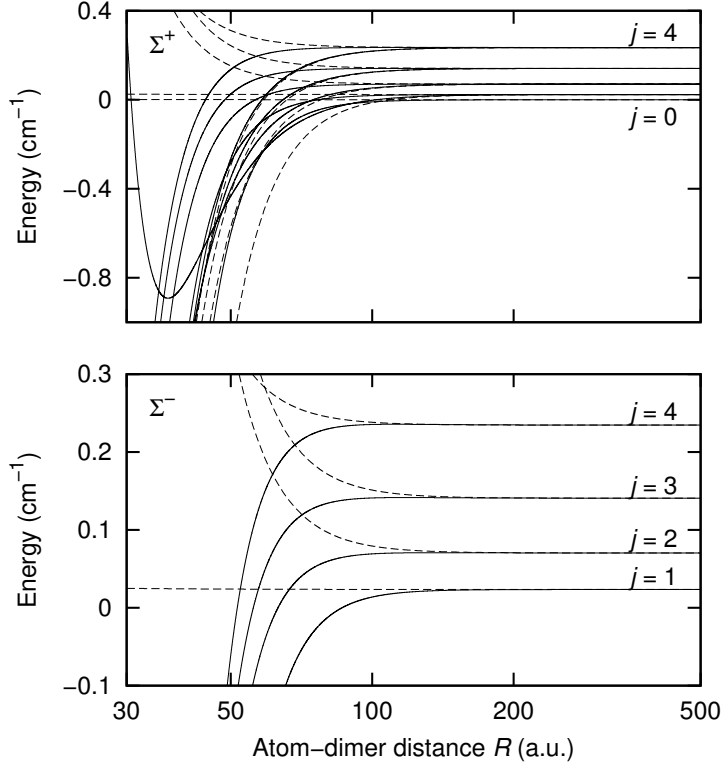}
\par\end{centering}
\caption{Long-range potential energy curves $B_{0}j(j+1) + C_{5}/R^{5} + C_{6}/R^{6}$ (full lines), and $B_{0}j(j+1) + C_{5}/R^{5}$ (dashed lines) as functions of the atom-dimer distance $R$ (in logarithmic scale), for the $\Sigma^+$ and $\Sigma^-$ symmetries, representing the interaction between Cs($6^2P$) and Cs$_2(X^1\Sigma_g^+, v_d=0, j)$ for $j=0$ to 4. The related $C_5$ and $C_6$ coefficients are given in Table \ref{tab:LR-C5-C6}.}
\label{fig:C56-Sigma}
\end{figure}

\begin{figure}
\begin{centering}
\includegraphics[width=0.6\textwidth]{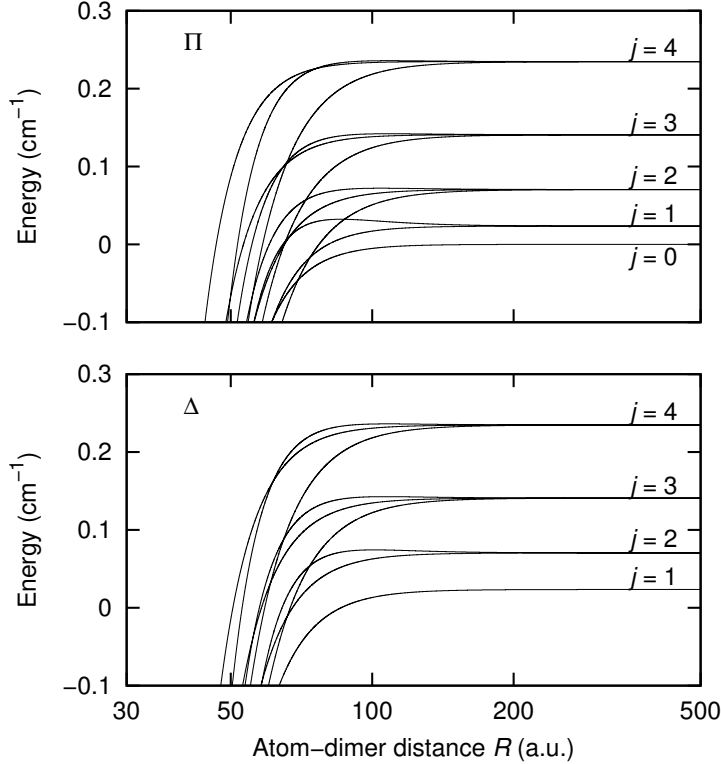}
\par\end{centering}
\caption{Same as Figure \ref{fig:C56-Sigma} for $B_{0}j(j+1) + C_{5}/R^{5} + C_{6}/R^{6}$ curves of $\Pi$ and $\Delta$ symmetries.}
\label{fig:C56-Pi-Delta}
\end{figure}

\begin{figure}
\begin{centering}
\includegraphics[width=0.6\textwidth]{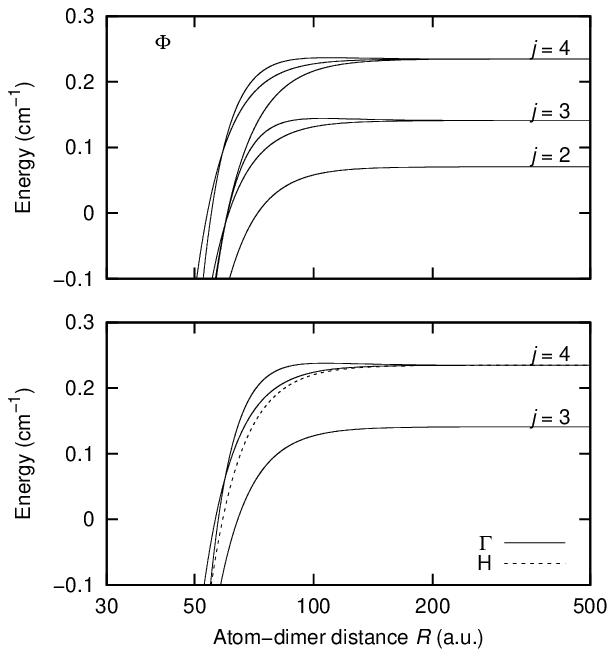}
\par\end{centering}
\caption{Same as Figure \ref{fig:C56-Pi-Delta} for the for the $\Phi$ symmetry (upper panel), and for $\Gamma$ (solid line) and H (dashed line) symmetries (lower panel).}
\label{fig:C56-Phi-Gamma-H}
\end{figure}

The potential energy curves for the long-range interaction between Cs($6^2P$) and Cs$_2(X^1\Sigma_g^+, v_d=0, j)$ are displayed in Figs. \ref{fig:C56-Sigma}, \ref{fig:C56-Pi-Delta}, and \ref{fig:C56-Phi-Gamma-H}.
For the $\Sigma$ symmetries (see Fig. \ref{fig:C56-Sigma}), the curves calculated in Paper I (only containing the $C_5$ term) are also shown. The related $C_5$ (from paper I) and $C_6$ coefficients are listed in Table \ref{tab:LR-C5-C6}.

Due to large and negative $C_6$ values, most potential curves are attractive. When the $C_5$ coefficients are positive, tiny potential barriers are visible. For instance, the highest barrier is found for $\Pi$ symmetry with a height of about  0.1~cm$^{-1}$ in the curve correlated to the $j=1$ rotational state. In paper I, we found that the low-$R$ limit of our perturbation analysis $R_m=102$~a.u. was imposed by the crossings between curves with different values of $j$. We see on Fig. \ref{fig:C56-Sigma}, that the typical position of these crossings is only slightly modified by the $C_6$ contribution. However, it is worth noting that the crossing range coincides with the one in which the second-order contribution competes with the first-order contribution. As a consequence, the non-adiabatic couplings emerging in the crossing region, evoked in paper I, should contain both first-order and second-order terms. The resulting couplings will mix states characterized by a given symmetry ($\Sigma^{\pm},\,\Pi,...$) and a given $j$, with states characterized by the same symmetry and by $j'=j\pm2$.

Note that for the $\Sigma^+$ symmetry and $j=1$, one see a long-range potential well, due to a negative $C_5$ coefficient (-1674~a.u.) and a positive $C_6$ coefficient (51249~a.u.) (see Fig \ref{fig:C56-Sigma}). This well is 0.9-cm$^{-1}$ deep, and its minimum is located at $R=37$ a.u., slightly below the LeRoy radius ($\sim 45~$a.u.). We can expect that this long-range well will indeed exist, but that it will be modified by non-adiabatic couplings discussed above and by the electronic exchange. Therefore, its precise characterization requires quantum-chemical calculations.

\section{Conclusions}
\label{sec:conclusion}

Using a second-order perturbation approach, we have shown that the long-range interaction between an excited Cs($6^2P$) atom and a Cs$_2(X^1\Sigma_g^+, v_d=0, j)$ molecule is significantly modified compared to the first-order quadrupole-quadrupole interaction. Most potential curves exhibit an attractive behavior, except in a few cases where either a low potential barrier or a long-range well is visible. The barriers could prevent the collision to occur in the ultracold domain. The validity of the present approach is limited to a range of distances well beyond the radius defined by the conventional LeRoy criterion, due to the low energy spacings between molecular rotational levels. This is actually the main difference compared to atom-atom long-range interaction, which completely changes the physical conditions for PA of an atom with a molecule. The next step in the theory is to include non-adiabatic couplings between curves related to different rotational levels, which is currently in progress in our group. They will induce avoided crossings which may generate a complex collisional dynamics.

While the present theory is general, it is difficult to predict if this situation is similar for all alkali systems, either homonuclear or heteronuclear, as it strongly depends on the balance between the influence of the various parameters, namely the quadrupole moment of the atom and the dimer, and their dipole polarizability. As already stressed in paper I, other effects should be taken in account, like atomic spin-orbit interaction. The fine structure splitting in Cesium (554.1~cm$^{-1}$) is much larger than the rotational energy of Cs$_{2}$, so that this effect it will not dramatically change our description. However, it is important to stress that the above treatment was developed in the basis of the $LS$ coupling case. The appropriate inclusion of the hyperfine structure will add to the complexity of the potential curves of Figs. \ref{fig:C56-Sigma}, \ref{fig:C56-Pi-Delta}, and \ref{fig:C56-Phi-Gamma-H}, even it will not modify the main conclusion of the present study: the photoassociation of a ground state $X^1\Sigma_g^+$ alkali-metal dimer molecule with a ground state $nS_{1/2}$ alkali-metal atom is generally possible by exciting the dimer-atom system with a laser field red-detuned  from the $nS_{1/2}\to nP_{1/2,3/2}$ atomic transition.

\section*{Acknowledgments}

We thank M. Aymar for kindly providing us with Cs$_2$ potential curves and transition dipole moments. This work was done with the support of Triangle de la Physique under contract 2008-007T-QCCM (Quantum Control of Cold Molecules),  and National Science Foundation under grant PHY-0855622.

%

\end{document}